# Investigating the Reliability in Three RAID Storage Models and Effect of Ordering Replicas on Disks

Leila Namvari-Tazehkand[a], Saeid Pashazadeh[b]

[a,b]Faculty of Electrical and Computer Engineering, University of Tabriz, Tabriz,
East Azerbaijan, Iran
Pashazadeh@tabrizu.ac.ir

**Abstract**
One of the most important parts of cloud computing is storage devices, and Redundant Array of Independent Disks (RAID) systems are well known and frequently used storage devices. With the increasing production of data in cloud environments, we need high-reliable storage, given the importance of data. RAID system's reliability analysis is of particular significance in the area of cloud storage. Generally, data redundancy is used to create fault tolerance and increases the reliability of storage. This study has considered three examples of the simple RAID storage models and analyzed their reliability. All of which have a replication factor of two and have the same number of disks and reliabilities. The only difference is the model that they used for generating the redundancy. To compare these three models' reliability, we examined the degree of fault tolerance (FT) and calculated the models' reliability using the reliability block diagram (RBD). In this paper, the effect of redundancy's type and the blocks' arrangement on the system's reliability was investigated.

## 1. Introduction
Many companies have provided cloud storage platforms such as Dropbox, Google Drive, and Amazon EC2, which allow users to enjoy on-demand, anytime, anywhere services from a shared pool of configurable resources in the cloud [1]. In cloud environments, RAID systems are used to store data. The RAID system is one of the most important hardware tools for storing data even at national and global levels. Today, with the expanding use of cloud computing and data storage, the importance of storage systems is increasingly being considered. In a system with thousands of hard disks, disk failure usually occurs, but multi-disk failures are also common. To avoid losing data due to the failure of the disks, there must be a way to think. Therefore, data reliability is one of the main concerns in RAID systems [2]. To have a high level of reliability and fault tolerance in data storages the redundancy method is used. The replica block and parity code block[1] two major styles are redundancy which is used to recover data from failed disks [2,3]. Specialists in the field of data storage software RAID models have introduced a lot which is in each of them for redundancy of information used the replica block method or parity code block method or a combination of the two. Each of the redundancy methods, have their own advantages and disadvantages. In [4] once the data itself and one-time replica block is saved for maintenance of fault tolerance. In [5] in addition to the data itself saved, double replica block and one-time parity code block are saved. In [6,7] in addition to the data itself saved, double replica block is saved.

## 2. Materials and methods
Many studies have been conducted on the redundancy and reliability in RAID systems [1-10]. Table 1 shows the advantages and disadvantages of the two redundancy methods.

---
[1] Parity code is a simple way to add redundancy block by using Xor of two or more blocks.

Table 1. Comparison of redundancy methods [3,8].

|  | Replica block | Parity code block |
|---|---|---|
| Advantages | More Read and write speed | Less coding time |
|  | Less complexity | Less storage space |
| Disadvantages | More coding time | Less read and write speed |
|  | More storage space | More complexity |

### 2.1. Proposed method

In this paper, three examples of the simple known RAID storage models were introduced to compare the coverage of fault tolerance and reliability for various RAID models. The blocks storage in these models are as follows:

- In the first model, once a data block (e.x. $B_0$) is stored, the second and third replicas of this block ($M_0$ and $M'_0$) are stored in different disks [4]. We name this replication method Replica-Replica (RR). Figure 1 shows a sample of the RR replication model.
- In the Parity code-Parity code (PP) model, once a data block (e.x. $B_0$) is stored, xor of this block with other blocks (e.x. $B_0 \oplus B_2$) are stored in different disks [6] as is shown in Fig. 2.
- In the Replica-Parity code (RP) model, once a data block (e.x. $B_0$) is stored, the replica of this block ($M_0$) and the third replicas in the form of Parity code, which is the xor of that block with other blocks are stored (e.x. $B_0 \oplus B_3$) [1,9]. Figure 3 shows a sample of the RP replication model.

| $D_0$ | $D_1$ | $D_2$ | $D_3$ | $D_4$ |
|---|---|---|---|---|
| $B_0$ | $B_1$ | $B_2$ | $B_3$ | $B_4$ |
| $M_4$ | $M_0$ | $M_1$ | $M_2$ | $M_3$ |
| $M'_3$ | $M'_4$ | $M'_0$ | $M'_1$ | $M'_2$ |

Figure 1. RR storage model.

| $D_0$ | $D_1$ | $D_2$ | $D_3$ | $D_4$ |
|---|---|---|---|---|
| $B_0$ | $B_1$ | $B_2$ | $B_3$ | $B_4$ |
| $B_1 \oplus B_2$ | $B_2 \oplus B_3$ | $B_3 \oplus B_4$ | $B_4 \oplus B_0$ | $B_0 \oplus B_1$ |
| $B_3 \oplus B_4$ | $B_0 \oplus B_4$ | $B_0 \oplus B_1$ | $B_1 \oplus B_2$ | $B_2 \oplus B_3$ |

Figure 2. PP1 storage model.

| $D_0$ | $D_1$ | $D_2$ | $D_3$ | $D_4$ |
|---|---|---|---|---|
| $B_0$ | $B_1$ | $B_2$ | $B_3$ | $B_4$ |
| $M_4$ | $M_0$ | $M_1$ | $M_2$ | $M_3$ |
| $B_0 \oplus B_2$ | $B_1 \oplus B_3$ | $B_2 \oplus B_4$ | $B_3 \oplus B_0$ | $B_4 \oplus B_1$ |

Figure 3. RP1 storage model.

### 2.1.1 The effect of ordering on degree of fault tolerance

We see that with the same conditions and assumptions, by changing the ordering model of blocks' contents of the PP1 and RP1 models, the degree of fault tolerance changes. Let name the new arrangements of the PP1 and the RP1 as PP2 and RP2, shown in Figures 4 and 5, respectively.

| $D_0$ | $D_1$ | $D_2$ | $D_3$ | $D_4$ |
|---|---|---|---|---|
| $B_0$ | $B_1$ | $B_2$ | $B_3$ | $B_4$ |
| $B_0 \oplus B_2$ | $B_1 \oplus B_3$ | $B_2 \oplus B_4$ | $B_3 \oplus B_0$ | $B_4 \oplus B_1$ |
| $B_3 \oplus B_4$ | $B_0 \oplus B_4$ | $B_0 \oplus B_1$ | $B_1 \oplus B_2$ | $B_2 \oplus B_3$ |

Figure 4. PP2 storage model.

| $D_0$ | $D_1$ | $D_2$ | $D_3$ | $D_4$ |
|---|---|---|---|---|
| $B_0$ | $B_1$ | $B_2$ | $B_3$ | $B_4$ |
| $M_4$ | $M_0$ | $M_1$ | $M_2$ | $M_3$ |
| $B_1 \oplus B_2$ | $B_2 \oplus B_3$ | $B_3 \oplus B_4$ | $B_4 \oplus B_0$ | $B_0 \oplus B_1$ |

Figure 5. RP2 storage model.

### 2.2. Analysis of degree of fault tolerance

According to Fig. 1, it can be seen in the RR model, the $D_2$, $D_3$, and $D_4$ will fail and the $D_0$ and $D_1$ are healthy. The $B_0$, $B_1$, $M_3$ and $M_4$ can be retrieved, but the $B_2$ can't be retrieved. So that the $D_1$ and $D_3$ and $D_4$ fail and the $D_0$ and $D_2$ are healthy. All blocks can be retrieved. According to Fig. 2, it can be seen in the PP1 model. the $D_2$, $D_3$, and $D_4$ have failed and the $D_0$ and $D_1$ are healthy. The $B_0$ and $B_1$ exist, and the $B_1$ xor with $B_1 \oplus B_2$, the $B_2$ can be retrieved and the $B_2$ xor with $B_2 \oplus B_3$, the $B_3$ can be retrieved, and the $B_3$ xor with $B_3 \oplus B_4$, the $B_4$ can be retrieved as a result all blocks can be retrieved. The degree of fault tolerance of all models can be achieved in the same way. By examining the different failure of the disks in all three models, we find that those all have a second-degree fault tolerance. That is, in the event of a simultaneous failure of two disks, the system can run with three disks. Table 2. shows the results of coverage of fault tolerance for RR and PP1 and RP1 models.

Table 2. Third degree fault tolerance (FT=3)

| Active Disks | RR | PP1 | RP1 |
|---|---|---|---|
| $D_0 D_1$ | ✗ | ✓ | ✓ |
| $D_0 D_2$ | ✓ | ✗ | ✗ |
| $D_0 D_3$ | ✓ | ✗ | ✗ |
| $D_0 D_4$ | ✗ | ✓ | ✓ |
| $D_1 D_2$ | ✗ | ✓ | ✓ |
| $D_1 D_3$ | ✓ | ✗ | ✗ |
| $D_1 D_4$ | ✓ | ✗ | ✗ |
| $D_2 D_3$ | ✗ | ✓ | ✓ |
| $D_2 D_4$ | ✓ | ✗ | ✗ |
| $D_3 D_4$ | ✗ | ✓ | ✓ |

Table 3. Shows the results of the fault tolerance of second-degree evaluation for the RR model and two versions of the proposed PP and RP models. Table 4. Shows the degree of fault tolerance for both PP2 and RP2 models. Any change in the RR configuration model will not improve its fault tolerance of third-degree.

Table 3. Second degree fault tolerance (FT=2)

| Active Disks | RR | PP1 PP2 | RP1 RP2 |
|---|---|---|---|
| $D_0 D_1 D_2$ | ✓ | ✓ | ✓ |
| $D_0 D_1 D_3$ | ✓ | ✓ | ✓ |
| $D_0 D_1 D_4$ | ✓ | ✓ | ✓ |
| $D_0 D_2 D_3$ | ✓ | ✓ | ✓ |
| $D_0 D_2 D_4$ | ✓ | ✓ | ✓ |
| $D_0 D_3 D_4$ | ✓ | ✓ | ✓ |
| $D_1 D_2 D_3$ | ✓ | ✓ | ✓ |
| $D_1 D_2 D_4$ | ✓ | ✓ | ✓ |
| $D_1 D_3 D_4$ | ✓ | ✓ | ✓ |
| $D_2 D_3 D_4$ | ✓ | ✓ | ✓ |

Table 4. Third degree fault tolerance (FT=3)

| Active Disks | PP2 | RP2 |
|---|---|---|
| $D_0 D_1$ | ✓ | ✓ |
| $D_0 D_2$ | ✓ | ✓ |
| $D_0 D_3$ | ✓ | ✓ |
| $D_0 D_4$ | ✓ | ✓ |
| $D_1 D_2$ | ✓ | ✓ |
| $D_1 D_3$ | ✓ | ✓ |
| $D_1 D_4$ | ✓ | ✓ |
| $D_2 D_3$ | ✓ | ✓ |
| $D_2 D_4$ | ✓ | ✓ |
| $D_3 D_4$ | ✓ | ✓ |

**2.3. Evaluation the reliability of model by means reliability block diagram**

Famous structures of reliability block diagram (RBD) of the systems have several types, most notably the following [10]:

- Series structure
- Parallel structure
- Series - parallel structure
- Parallel - series structure
- K-out-of-N structure (KooN)
- Non series – non parallel structure (bridges)

RAID storage models follow a parallel-series structure and its reliability is calculated according to relation (1). In the following relationship $i$ represents the number of parallel paths and $\rho_i(x)$ representing the series items available in each direction [10].

$$R_{\text{System}} = 1 - \prod_{i=1}^{n}(1 - \rho_i(x)) \qquad (1)$$

In the previous section we saw that the three models had second-degree of fault tolerance. Figure 6 shows the RBD of all models.

## 3. Results and discussions

We draw the RBDs of the previous section on the Windows 8.1 operating system using the SHARPE software and obtain the corresponding Reliability charts. For all models, we get the reliability of 10000 hours.

In Fig. 7-(a), we observe that the reliability of all models for fault tolerance of third degree is greater than their reliability for fault tolerance of second degree. In Fig. 7-(b), we see that the reliability of the PP1 and RP1 models for fault tolerance of third degree is greater than the reliability of RR for fault tolerance of second degree.

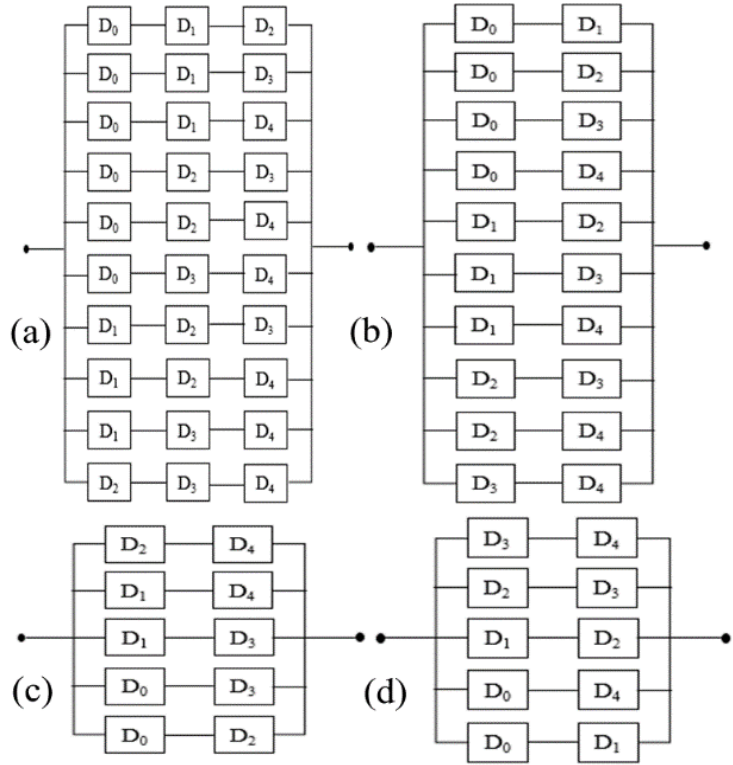

Figure 6: The RBD of (a) RR, PP and RP for ($FT = 2$), (b) PP2 and RP2 for ($FT = 3$), (c) RR for ($FT = 3$), and (d) PP1 and RP1 for (FT=3).

In Fig. 7-(c), we see that the reliability of the PP2 and RP2 models for fault tolerance of third degree is greater than RR's reliability for fault tolerance of third degree. Also, we see in Fig. 7-(d) that when the location of blocks in the PP1 and RP1 models changes, the reliability level has increased.

## 4. Conclusion

In this paper, with the aim of comparing the coverage of fault tolerance and reliability for various RAID models, three examples of the simplest RAID storage models were introduced, the degree of fault tolerance in the models was investigated, and with the aid of a reliability block diagram, by using of SHARPE software, the reliability of the models was checked and according to the results, all models, due to the repeat factor of two, are sure to have two degree fault tolerance. The RR and PP1 and RP1 models do not have third degree fault tolerance in all cases, but by changing the ordering of the blocks, PP2 and RP2 are 100% times have third degree fault tolerance. It can be concluded that for to have third degree fault tolerance, PP2 and RP2 have the same reliability, but RR has less reliability than the first two. In the proposed software RAIDs, you can combine a replica block and parity code block, in addition to the optimal use of disk space and reducing the complexity of data retrieval to provide reliability was relatively good.

Simulation results show that the replicas' new proposed ordering increases the RAID system's reliability compared to the previous ordering schema. Results show that appropriate placement of blocks in disks can improve the reliability of the RAID system in addition to the optimal use of disk space and reducing data retrieval complexity. This improvement in reliability increases with the increasing degree of Fault-tolerance.

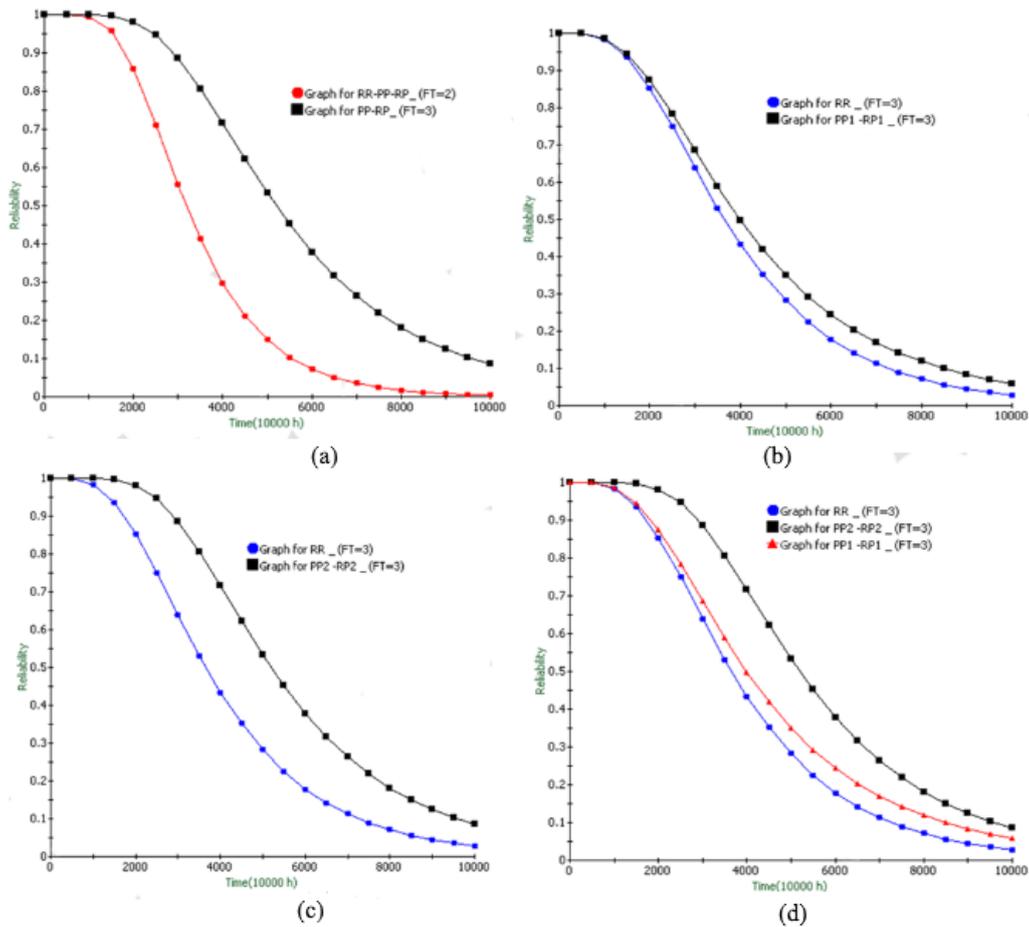

Figure 7: The reliability of (a) all models for ($FT = 2$) and ($FT = 3$), (b) RR, PP1, RP1 models for ($FT = 3$), (c) RR, PP2, RP2 models for ($FT = 3$), and (d) RR, PP1, RP1, PP2, RP2 ($FT = 3$).